\documentstyle[aps]{revtex}

\begin{document}
\draft
\title{Comparative Noninformativities of Quantum Priors
Based on Monotone Metrics}
\author{Paul B. Slater}
\address{ISBER,
University of California, Santa Barbara, CA 93106-2150\\
e-mail: slater@itp.ucsb.edu,
FAX: (805) 893-2790}

\date{\today}
\maketitle

\begin{abstract}
We consider a number of {\it prior}
 probability distributions of particular interest, all being defined on the
three-dimensional convex set of two-level quantum
systems. Each distribution  is  --- following recent work of
Petz and Sudar --- taken to be proportional to the volume element of a
{\it monotone} metric
on that Riemannian manifold.
 We apply an entropy-based test (a variant of one
 recently developed by Clarke) to determine which of two  priors is
more {\it noninformative} in nature. This involves
converting them to
 {\it posterior} probability distributions based on some set of
 hypothesized outcomes of measurements
of the quantum system in question. It is, then,  ascertained whether
or not the relative entropy (Kullback-Leibler statistic) between a pair
of priors
 {\it increases}
 or {\it decreases} when one of them is exchanged with its
corresponding
posterior. The findings lead us to assert
that the maximal
 monotone metric yields the {\it most} noninformative prior
distribution and the minimal
monotone (that is, the Bures) metric, the {\it least}.
Our conclusions both agree and disagree, in certain respects, with ones
recently reached by Hall, who relied upon a less specific test criterion
than our entropy-based one.

\end{abstract}

\vspace{.1in}


\hspace{1.4cm} Keywords: noninformative priors, Bayesian inference,
 spin-${1 \over 2}$
systems, monotone metrics,

\hspace{3.2cm} relative entropy, posterior distributions, quantum measurements

\vspace{.5cm}

\section{Introduction}
At the outset of this letter, let us note that subsequent to an earlier
version of it (quant-ph/9703012),
Hall communicated a study \cite{hall}, having quite similar (though
perhaps more clearly articulated)
objectives. Nevertheless, in terms of methodologies and conclusions,
 the two studies appear to differ significantly. 
We shall indicate these interesting points of agreement and disagreement.

Let us
reiterate the fundamental question motivating  Hall, which he
 states at the beginning of his 
presentation: ``what statistical ensemble corresponds to {\it minimal} prior
knowledge about a quantum system? Such an ensemble may be identified as the
most {\it random} ensemble of possible states of the system. It would
provide, for example, a natural benchmark for assessing how ``random''
a given evolution process is; a worst-case scenario for general schemes for
extracting information  about the system; and a natural unbiased measure over
the set of possible states of the system (which would allow one to calculate,
e. g., the average effectiveness of a general scheme for distinguishing
between quantum states)''.

In our work presented below,
 we have relied  upon a specific entropy-based
 test (related to one recently developed by
 Clarke 
\cite{clarke96})  to address
these issues Hall has raised. The conclusions of Hall themselves,
 on the other hand, 
 rest on the less specific (and apparently, from our point of view,
 not determinative)
proposition that ``maximal randomness corresponds to an ensemble with
maximal symmetry.'' In fact, we will argue that Hall's minimal-knowledge
ensemble (based on the Bures metric \cite{hub}) is not truly minimal.
 However, we do agree with Hall that this particular
ensemble is superior in information terms to that associated with a uniform
distribution (as previously employed by Larson and Dukes \cite{larson})
 over the convex set (Bloch sphere \cite{BRAUN}) of two-level quantum systems.

 Brody and Meister \cite{brody} recently studied the problem of deciding
between two {\it a priori} possible
two-level quantum mechanical pure states.
 They assert that ``if prior knowledge is not
available, one can still employ the Bayesian approach, using a noninformative
prior. However, the analysis of such cases is beyond the scope of the present
Letter.'' In this study, we {\it do} consider the problem of determining an
appropriate noninformative prior --- for the more general situation,
in which advance information regarding the degree of purity
of the unknown system is lacking.
(``In a real situation one can never design a preparator such that it
produces an ensemble of identical pure states. What usually happens is
that the ensemble consists of a set of pure states each of which is
represented in the ensemble with a certain probability'' \cite{derka1}.)

As noted, we adapt recent work of Clarke \cite{clarke96} to provide us with
an operational criterion for deciding which of two priors
is more noninformative in nature.
We apply our  test to a number of priors, all of
which (with the exception of the uniform distribution of Larson and Dukes
\cite{larson})
 are obtained by normalizing the volume elements of monotone metrics
\cite{petz2,petzsud,petz1}.
We find that the maximal monotone metric (of the left logarithmic
derivative --- which does not exist for pure states)
 yields the {\it most} noninformative prior of those examined,
 but only
if we first rule out the possibility that the unknown two-level system is in a
pure or nearly pure state.
(Jones \cite{jones} has
 considered the somewhat opposite situation, in which the
unknown system is {\it definitely} in a pure state.)
We also find --- in strong contrast to Hall \cite{hall} --- that
 the minimal monotone metric (of the symmetric logarithmic
derivative) yields the {\it least} noninformative prior of those examined
(other than the uniform one of Larson and Dukes \cite{larson}).
It also appears that our (Bayesian)
 notion of noninformativity is equivalent to the (classical) one
of Petz and Toth \cite[p. 215]{petz3} in their work comparing 
 (Cram\'{e}r-Rao-type) lower
bounds for the variances/covariances of unbiased estimators of the
parameters of quantum systems.

In (classical/nonquantum) Bayesian theory \cite{bern}, in the absence of
any information regarding the specific values of the parameters of a family
of probability distributions, one uses a 
noninformative (Jeffreys)
 prior (cf. \cite{bal1,bal2}). This is taken to be proportional to
 the volume element of the {\it unique} monotone (Fisher
information) metric on the Riemannian manifold formed by the family.
``An infinitesimal statistical distance has to be monotone under stochastic
mappings'' \cite[p. 786]{petz2}.
 Petz and Sudar \cite{petzsud} --- building upon work of Morozova and Chentsov \cite{chentmor} --- have recently
shown that, in the quantum/noncommutative case, there is not a single
 monotone metric, but rather a nondenumerable number of such metrics.
 Each corresponds
to an operator
monotone function $f(t)$ satisfying the normalization and symmetrization
conditions,
$f(1)=1$, $f(t)=t f(t^{-1})$.
(A function $f: \Re^+ \rightarrow \Re$ is called operator
monotone if the relation $0 \leq K \leq H$,
meaning that $H-K$ is semipositive definite, implies $0 \leq f(K) \leq f(H)$,
for any matrices $K$ and $H$ of arbitrary order \cite{donog}.)
``Therefore, more than one privileged metric shows up in
quantum mechanics. The exact clarification of this point
requires and is worth further studies'' \cite[p. 2672]{petzsud}.
This Letter seeks to contribute to such a clarification (cf. \cite{kratt}).
 We, thus, restrict our
considerations to priors which are proportional to volume elements
of monotone metrics. Although this still leaves a nondenumerable
number of
candidate priors, our results indicate that certain ones (in
particular, that obtained from the maximal monotone metric) can be
distinguished for their information-theoretic properties.

\section{Analysis}
\subsection{Two  prior probability distributions}

We begin our analysis by examining two
 monotone metrics of particular interest \cite{petzsud} --- the Kubo-Mori
metric (given by $f(t) = (t-1)/ {\log t}$) and the minimal
monotone SLD (symmetric logarithmic
derivative) ``Bures-type'' \cite{braun} metric (given by $f(t)=(1+t)/2$).
In recent papers \cite{slat} (cf. \cite{kratt,slat2}),
the author has proposed and analyzed the use of a prior
probability distribution,
\begin{equation} \label{eq:1}
(1-x^2-y^2-z^2)^{-1/2}/{\pi}^2.
\end{equation}
This is the normalized form of the volume element of
 the SLD-metric over the Bloch sphere
\cite{braun2} --- the unit ball in three-space, comprised
of the points $x^2+y^2+z^2 \leq 1$ --- of
$2\times 2$ density matrices,
\begin{equation} \label{eq:3}
 {1 \over 2}
\pmatrix{\ 1+z&x-iy\cr
         x+iy&1-z\cr}.
\end{equation}
In spherical coordinates ($x=r \cos \phi \sin \theta,y=r \sin \phi \sin
\theta, z =r \cos \theta$), (\ref{eq:1}) takes the form,
\begin{equation} \label{eq:2}
p_{SLD}(r,\theta,\phi) = r^2 (1-r^2)^{-1/2} \sin \theta/{\pi}^2.
\end{equation}
On the other hand, use of the Kubo-Mori metric yields a prior
probability distribution,
\begin{equation} \label{eq:4}
p_{KM}(r,\theta,\phi) = r (1-r^2)^{-1/2} \log[(1+r)/(1-r)] \sin \theta/{4 \pi^2}.
\end{equation}
The distributions
 (\ref{eq:2}) and (\ref{eq:4}) are obtained by substituting the
corresponding operator monotone functions given above into
the formula \cite[eq, 3.17]{petzsud},
\begin{equation} \label{eq:4-5}
{r^2 (1-r^2)^{-1/2} (1+r)^{-1} \sin \theta} /f[(1-r)/(1+r)],
\end{equation}
and, then, normalizing.
Both $p_{KM}$ and $p_{SLD}$ are monotonically increasing with $r$,
with $p_{KM}$ assigning greater probability to systems that are nearly pure
($r > .957504$) and, compensatingly less, to relatively mixed systems
($r < .957504$). 

We, first,  compare the suitabilities
of $p_{SLD}$ and $p_{KM}$ as possible
noninformative or ``reference'' priors \cite{bern} for the quantum inference or
estimation of an unknown
two-level system.
(Appropriate {\it informative} priors should, of course, be used if
specific knowledge regarding the parameters of the system is available
\cite{brody}.)
 We
modify (as explained below) --- in any case, doing so apparently, at least in
our context, leads to no substantive differences --- a
 general line of reasoning recently elaborated upon
by Clarke \cite{clarke96}.
We note, in this regard,
 that the relative entropy (Kullback-Leibler number or information
gain or directed divergence)
\cite{bern} of $p_{SLD}$ with respect to $p_{KM}$, that is,
\begin{equation} \label{eq:5}
D(p_{SLD} \parallel p_{KM}) = \int_{0}^{1} \int_{0}^{\pi} \int_{0}^{2 \pi}
 p_{SLD} \log[p_{SLD}/p_{KM}] d \phi d \theta dr
\end{equation}
(the natural logarithm is understood throughout this communication)
is .0891523 ``nats'' (1 nat equals ${1 \over {\log 2}} \approx 1.4227$ bits),
 while $D(p_{KM} \parallel p_{SLD})$ is .0975976 nats.
It proves possible to reduce the former statistic
by incorporating certain information into our considerations,
but not the latter. This leads us to the conclusion that
$p_{KM}$ is more noninformative than $p_{SLD}$.

After some initial numerical experimentation, we were led to
 assume that six spin measurements had been performed --- two in the $X-$, two in the $Y-$, and two
in the $Z$-direction --- using six replicas of a spin-1/2 system and that for each of these pairs we obtained one ``up''
and one ``down''. 
(It would be of interest to conduct a parallel series of analyses
to that reported below, based on what have been found to be
{\it optimal} sets of measurements \cite{derkaopt,latorre}.) Then --- applying
Bayes' Theorem \cite{bern} --- we converted $p_{SLD}$ and $p_{KM}$ to
posterior distributions by multiplying them by the {\it likelihood}
of such a set of six outcomes
\cite{slat,jones},
\begin{equation} \label{eq:6}
(1-x^2)(1-y^2)(1-z^2)/64 =[(1-x)/2][(1+x)/2][(1-y)/2][(1+y)/2][(1-z)/2][(1+z)/2],
\end{equation}
and normalizing the resulting
 products over the Bloch sphere. (The normalization
factors are $64 \times 192/71$ in the $SLD$-case and
$ 64 \times 19600/6047$ in the $KM$-case.)
The relative entropy of $p_{SLD}$ with respect to the $KM$-posterior
 is, then, {\it reduced}, as a result of
 the {\it added} information, from .0891523 to .0720681.
On the other hand, the relative entropy of $p_{KM}$ with respect to the
$SLD$-posterior is {\it increased} dramatically
 from .0975976 to .457259.
Paraphrasing Clarke \cite[p. 173]{clarke96}, ``[$p_{SLD}]$ is already more
informative than [$p_{KM}$], so we cannot make it less informative
by adding information''. However, if we were to replace the likelihood
(\ref{eq:6}) by its square --- that is, in effect, assume twelve measurements,
giving {\it two} ``ups'' and {\it two} ``downs'' in each of the three
mutually orthogonal directions --- then the relative entropy of $p_{SLD}$
with respect to the corresponding
 revised or updated $KM$-posterior would not further
decrease from .0720681, but would increase to .334699. 
Thus, the informativity of $p_{SLD}$ with respect to $p_{KM}$ is limited,
in this manner.
(In an approximate sense, then, the information contained in $p_{SLD}$ can
be described as that in $p_{KM}$ with the addition of that gained
by knowledge of the outcomes of the six measurements.)

If we had conformed strictly to the line of argument of Clarke \cite{clarke96},
we would have exchanged the positions of the priors and posteriors
 in the relative entropy statistics reported above.
Nevertheless, it seems rather evident that --- in the context of
the present study --- we would reach the same fundamental conclusions if we
had done so.
 For example (again, based on the same six measurement outcomes), the relative
entropy of the $KM$-posterior with respect to $p_{SLD}$ is .0603743
(cf. .0720681) and that of
the $SLD$-posterior with respect to $p_{KM}$ is .399442
(cf. .457259).
(Even though, relative entropy ``is not a true distance
between distributions since it is not symmetric
and does not satisfy the triangle inequality...it
is often useful to think of relative entropy
as a `distance' between distributions''
\cite[p.18]{cover}). Our initial rationale
 for not simply following
 Clarke's scheme was based, among other things, on cetain preliminary
evidence that
computations would be considerably simplified if we averaged
the logarithms of the ratios of the two probability distributions
in question with respect to
(spherically-symmetric) priors and not (asymmetric) posteriors.
Since the supports of the priors and posteriors are essentially identical
here, that is the Bloch sphere --- except
 for possibly the six isolated points, $(\pm 1,0,0),(0,\pm 1,0)$ and
$(0,0, \pm 1)$, having measure zero --- our
 variation will not lead to divergent integrals.
However,
since the support of a posterior, in general, may be a measurably
smaller subset
of the support of the corresponding prior (due to the likelihood being null),
the scheme of Clarke and not our variation should be followed, as a rule.
(For those priors studied here which are defined over the entire Bloch
sphere, the likelihood (\ref{eq:6})
is zero at the six mentioned points. However, all these priors --- except
 for the uniform distribution ($p_{LD}$) --- are infinite at
those points also.) Later developments appeared to, in fact, indicate
that we would not have paid a significant computational penalty in fully
adhering to the approach of Clarke, {\it ab initio}.

Let us present two additional information-theoretic statistics
consistent with the proposition that $p_{KM}$ is more noninformative
than $p_{SLD}$. The information gain of the $KM$-posterior
(based on the same
 hypothetical six observations) with respect to $p_{KM}$ itself
is .151575, while the analogous $SLD$-result  is less,
$ {4693 \over 1420} + \log {3 \over 71} \approx .140862$.
Also, a {\it single} spin measurement yields an information gain of
.157404 with respect to $p_{KM}$, but less
 ($ {5 \over 6} - \log 2 \approx .140186$) with respect to $p_{SLD}$.
\subsection{Two additional prior probability distributions}

Using a recursively defined double sequence, Petz \cite[eq. 21]{petz1} has
arrived at the operator monotone function,
\begin{equation} \label{eq:7}
{f(t)} = {{2 (t-1)^2} \over {(1+t) (\log t)^2}},
\end{equation}
the operator mean of which had been found by Morozova and Chentsov
\cite{chentmor}.
Normalizing, using numerical methods,
 the volume element (\ref{eq:4-5}) of the associated monotone
 metric,
we obtain the following prior probability distribution,
\begin{equation} \label{eq:8}
p_{MC}(r,\theta,\phi) = .00513299 (1-r^2)^{-1/2} ( \log [(1-r)/(1+r)])^2 \sin \theta.
\end{equation}
Now, $D(p_{KM} \parallel p_{MC}) = .112421$
 and $D(p_{MC} \parallel p_{KM}) = .117982$.
These statistics are transformed to .106655 and .482023, respectively,
when they are computed with respect to the
corresponding $MC$ and $KM$-posteriors, both
 based on the previously
 hypothesized set of six outcomes.
 So, we can assert --- by our rule --- that $p_{MC}$ is itself more
noninformative than $p_{KM}$.
(Strict adherence to the scheme of Clarke yields the corresponding statistics,
.0910048 and .452794, leading to the very
same conclusion. It appears, however, that
 our procedure yields somewhat larger statistics
than does Clarke's.) This is consistent with the earlier
result, in the sense that $p_{MC}$ assigns greater probability
to some set of nearly pure systems ($r > .9846$) than does $p_{KM}$.
Assuming that noninformativity is a transitive relation,
one would expect to find that $p_{MC}$ is also
 more noninformative than $p_{SLD}$.
This proves to be the case, as $D(p_{SLD} \parallel p_{MC}) = .388323$
and $D(p_{MC} \parallel p_{SLD}) =.445981$.
The posterior version of these statistics are, .186964 and .991175,
respectively, with $p_{MC}$ assigning greater probability than $p_{SLD}$
to systems with $r > .973932$.

Larson and Dukes \cite{larson}
 have utilized a uniform prior,
\begin{equation} \label{eq:9}
p_{LD}(r,\theta,\phi) = 3 r^2 \sin \theta /4 \pi,
\end{equation}
 over the Bloch sphere
of two-level quantum systems.
``The simplest prior which does not confine itself to pure states assigns
equal probability to equal infinitesimal volumes within the unit sphere
of the geometrical parameterization. (This is a physically
 reasonable prior, since the
 geometrical parameterization is metrically faithful'' \cite{larson}.
Hall \cite{hall} obtains the uniform distribution by two distinct arguments,
one based on randomly correlated ensembles, and the other on the
Hilbert-Schmidt metric.)

 Although $p_{LD}$ can be written as proportional to the
volume element (\ref{eq:4-5}), using $f(t)=(1+t)^{2}/ \sqrt t$,
this particular function is {\it not} operator monotone,
so $p_{LD}$ does not, in fact, correspond to a monotone metric.
 We can, nevertheless,
 examine its informative
properties.
We have that $D(p_{LD} \parallel p_{MC}) =1.07895$
and $D(p_{MC} \parallel p_{LD}) = 1.98719$.
These statistics are transformed to .559829 and 2.79851, respectively,
if one replaces the second probability distribution in each formula
with its posterior counterpart based on the set of six outcomes
previously hypothesized
 (\ref{eq:6}). These results are, then, consistent
with the earlier patterns, since $p_{MC}$ assigns greater probability
than $p_{LD}$
to systems with $r > .948724$.
By assuming a doubling or repeating of
 the set of six measurements --- involving
the squaring of the likelihood (\ref{eq:6}) --- we are able to
reduce the statistic .559829 still further, to .310686. while a tripling
(corresponding to eighteen measurements) yields less still, that is .307632.
However, use of a posterior based on twenty-four such measurements
results in .529577. So, it might be asserted that $p_{LD}$ is
{\it considerably more} informative than $p_{MC}$.

The variance of $z$, that is $\langle z^2 \rangle$, is .301762 for $p_{MC}$,
.277778 for $p_{KM}$, .25 for $p_{SLD}$ and .2 for $p_{LD}$,
thus, agreeing in order with the relative noninformativities of these
priors. In the relative ranking of $p_{SLD}$ and $p_{LD}$, we are in
full agreement with Hall \cite{hall}. Hall's argument that $p_{SLD}$ 
generates the maximally random ensemble is that ``the Bures metric for
a two-dimensional system corresponds to the surface of a unit 4-ball, i. e.,
to the maximally symmetric 3-dimensional space of positive curvature \ldots
This space is homogenous and isotropic, and hence the Bures metric does not
distinguish a preferred location or direction in the space of density
operators.'' Although we can not disagree with these statements, they do
not appear to be determinative in judging the relative noninformativity
of prior distributions over the two-level quantum systems.
\subsection{Two {\it truncated} prior probability distributions}
We now proceed to find two priors more noninformative than $p_{MC}$, but
only by imposing a restriction on the {\it a priori} possible two-level
systems --- that is, we must eliminate the possibility that the unknown
system under examination is either in a pure or ``nearly'' pure state.
We consider the three operator monotone functions \cite{petzsud,petz1},
\begin{equation} \label{2}
 {f(t)} ={{2 t^{(2-n) \over 2} (t-1)^n} \over {(1+t) (\log t)^n}}. \qquad n=0,1,2
\end{equation}
For $n = 0$ (corresponding to the {\it maximal} monotone metric [of the
left logarithmic derivative] \cite{petzsud,petz1}) and $n= 1$, the volume elements (\ref{eq:4-5}) are {\it improper}, that is {\it not}
normalizable over the Bloch sphere, while for $n=2$, the corresponding
volume element is proper or normalizable, corresponding to the operator
monotone function (\ref{eq:7}) and probability distribution $p_{MC}$,
that is (\ref{eq:8}).
 To directly compare the three
metrics based on (\ref{2}), we choose to normalize their volume elements over a three-dimensional
ball of radius $R = 1-10^{-10}$. We, consequently, obtain the three
probability distributions ($p_{n}$) over the so-truncated convex set
(not containing the pure [$r=1$] and nearly pure [$1>r>R$] states),
\begin{equation} \label{3}
p_{0} = .00000112542 r^2 (1-r^2)^{-3/2} \sin \theta,
\end{equation}
\begin{equation} \label{4}
p_{1} =.000569121 r (1-r^2)^{-1} \log [(1+r)/(1-r)] \sin \theta,
\end{equation}
\begin{equation} \label{5}
p_{2} = .00513611 (1-r^2)^{-1/2} {\log [(1-r)/(1+r)]}^{2} \sin \theta.
\end{equation}
We have that the relative entropy or information gain of $p_{0}$ with
respect to $p_{1}$, that is [cf. (\ref{eq:5})], 
\begin{equation}
D(p_{0} \parallel p_{1}) = \int_{0}^{R} \int_{0}^{\pi} \int_{0}^{2 \pi}
p_{0} \log [p_{0}/p_{1}] d \phi d \theta dr
\end{equation}
equals .867442 nats.
Also, $D(p_{0} \parallel p_{2})$ =5.76086,
$D(p_{1} \parallel p_{0})$ = 1.654,
$D(p_{1} \parallel p_{2})$ = 2.37198,
$D(p_{2} \parallel p_{0})$ =7.06816, and
$D(p_{2} \parallel p_{1})$ = 1.52109.

We assume now --- again following our general methodology --- that
 we have performed a two-level measurement
on each of six replicas of a two-level quantum system, two measurements
in each of three orthogonal ($X,Y,Z$) directions, and obtained a single
``up'' and a single ``down'' in each direction.
Multiplying the likelihood (\ref{eq:6}) of such an occurrence by $p_{0},p_{1}$ and $p_{2}$, in turn, and normalizing
the products over the truncated Bloch sphere (of radius $R=1-10^{-10}$) --- with normalization factors, 335.987, 327.546 and 249.378, respectively --- we
 obtain (in accordance with Bayes' Theorem \cite{bern}) three posterior distributions, which we will designate as
$P_{0},P_{1}$ and $P_{2}$, in the obvious fashion. Now, we have that
$D(p_{0} \parallel P_{1}) = 1.07576$,
$D(p_{0} \parallel P_{2}) = 6.24184$,
$D(p_{1} \parallel P_{0}) = 1.53564$,
$D(p_{1} \parallel P_{2}) = 2.55172$,
$D(p_{2} \parallel P_{0}) = 6.94979$,
and 
$D(p_{2} \parallel P_{1}) = 1.42817$.

Our variation of the
 criterion elaborated upon by Clarke \cite{clarke96} (in which
we have exchanged the positions of priors and posteriors in the relative
entropy statistics), then leads us
to conclude that $p_{0}$ is more noninformative than both $p_{1}$ and
$p_{2}$ and that $p_{1}$ is more noninformative than $p_{2}$.
According to our rule, $p_{i}$ is more noninformative than $p_{j}$ if
both
$D(p_{i} \parallel P_{j}) > D(p_{i} \parallel p_{j})$ and
$D(p_{j} \parallel P_{i}) < D(p_{j} \parallel p_{i})$.
For instance, for $i=0$ and $j=1$, we have that
$D(p_{0} \parallel P_{1}) = 1.07576 > D(p_{0} \parallel p_{1}) = .867442$
and
$D(p_{1} \parallel P_{0}) =1.53564 < D(p_{1} \parallel p_{0}) = 1.654$.
Thus, by adding information to $p_{0}$ --- in the form of the
six hypothesized measurement results --- we are able to more closely
approximate $p_{1}$, but apparently not {\it vice versa}.
It would be of interest to study the changes in the statistics given above as
$R \rightarrow 1$.
The ratio of $p_{0}$ to $p_{1}$ at 
$r = R = 1 - 10^{-10}$
 is 5.89521, while that of $p_{0}$
to $p_{2}$ is 1947.41, and that of $p_{1}$ to $p_{2}$ is 330.338.
These results are in accordance with those above, in that more noninformative priors were also found there to assign greater probability
to more nearly pure states.

\section{Discussion}

Petz and Toth \cite[p. 215]{petz3} found that the lower bound on
the covariance matrix of
unbiased estimators of parameters
 provided by use of the symmetric logarithmic derivative
or minimal montone metric
was ``more informative'' (that is, tighter) than the bound furnished by
the Kubo-Mori (Bogoliubov) monotone metric. This bound,
 in turn, was tighter than
that obtained from the maximal monotone metric. Their notion of informativity
 appears to be fully consistent with that derived here through an
apparently quite
different line of argument. We assert this
 since we have found that the noninformativity of
$p_{SLD}$ is less than that of $p_{KM}$, and that of $p_{KM}$ is less than
that of $p_{MC}$, while finally, the noninformativity of the truncated form
of $p_{MC}$, that is $p_{2}$, is less than that of $p_{0}$ (based on the
maximal monotone metric).

Derka, Buzek and Adam \cite{derka1} (cf. \cite{derka2}) approach the
problem of using Bayesian reasoning to reconstruct a two-level
quantum system that is possibly impure,
 by assuming that the unknown system is coupled to a second system
(which they, for convenience, take to be two-level in nature),
and that the pair of coupled systems is in a pure state.
They utilize the invariant integration measure on all possible such
(four-dimensional)
pure states. Hall \cite{hall} adopts a related approach in the first part
of his paper.
In contrast, the analysis here and in \cite{slat,slat2}
does not posit any such coupling of the unknown system, regarding it,
in effect, as independent or autonomous.

A most interesting question that remains to be formally
addressed is whether or not it is possible
to find two different sets of measurements, one of which leads to a
conclusion that a prior $p$ is more noninformative than another prior $q$,
while the other set leads to an opposite deduction. Of course, if
 such hypothetical sets were to exist, which we presume they do not,
 the validity of the general line of
 argument presented
 in this Letter would
be called into question.
Another line of investigation that would be  of interest to pursue would be the
determination of
 the particular set of measurements that {\it minimizes} the
relative entropy between the corresponding posterior form ($P$) of $p$ and $q$
itself (cf. \cite{jones}). Such a set of minimizing measurements could be said
to best
express the additional information contained in $q$,
above and beyond that in $p$. 
(Clarke \cite{clarke96} argues that one can, in general, find
a data set to minimize the relative entropy of the corresponding
posterior with respect to the ``informative'' prior in question.)

In summary, based on our analysis here, we would recommend for the Bayesian
inference of the parameters of an unknown two-level quantum system
the use of $p_{MC}$ as a prior or, preferably, some version (possibly
modifying our choice of $R$) of
 $p_{0}$, if one has {\it a priori}
knowledge that the unknown system is described by a polarization vector
of length $ \sqrt {x^2+y^2+z^2} \leq R  < 1$.
If we view $p_{0}$
 as a function of $R$, reexpress it in terms of
Cartesian coordinates, integrate over $z$, say, and take the limit
 $R \rightarrow 1$,
we obtain the {\it bivariate marginal} probability distribution over the
unit disk,
\begin{equation} \label{int}
(1-x^2-y^2)^{-1/2}/{2 \pi}.
\end{equation}
(The bivariate marginal distribution of $p_{SLD}$, on the other hand,
is the {\it uniform} one --- $1/ \pi$ --- over
the unit disk. In \cite{slat2} the distribution
 (\ref{int}) was obtained from the Jeffreys prior
for the family of trivariate normal distributions, with null mean
vectors, having the density matrices
(\ref{eq:3}) as their covariance matrices.)
Thus, if one is content to estimate only two of the three parameters
determining a two-level system, one can employ (\ref{int}) as a prior
and avoid having to rule out the possibility of a pure or nearly pure state
\cite{slat2}.
The distribution (\ref{int}) is precisely the standard
(classical/nonquantum) noninformative (Jeffreys) prior for the two-parameter
family of trinomial distributions with probabilities $x^2$, $y^2$
and $1-x^2-y^2$ \cite{bern,slat2}.
The {\it conditional} distribution over [-1,1] of $x$ in (\ref{int})
 --- given that $y=0$ --- is
the ``cosine distribution'' \cite{dariano}
\begin{equation} \label{int1}
(1-x^2)^{-1/2}/ \pi.
\end{equation}
(In the statistical literature this is termed the arc-sine distribution.)
It is the noninformative
(Jeffreys) prior
for the one-parameter family of binomial distributions
with probabilities $x^2$ and $1-x^2$
\cite{bern,slat2}.
The distributions (\ref{int}) and (\ref{int1}) can also be obtained
as conditional distributions of $p_{SLD}$ --- which itself is the Jeffreys
prior for the family of {\it quadrinomial} distributions with probabilities
$x^2, y^2, z^2$ and $1-x^2-y^2-z^2$.
(This corresponds to the geometry of a three-dimensional hemisphere,
as pointed out in \cite{hall,hub,braun2} (cf. \cite{kass}).)

For additional quantum applications of other (classical) work --- besides
\cite{clarke96} ---
 of Clarke (jointly
with Barron) \cite{clarkebarron1,clarkebarron2}, having relevance to
comparative properties of the volume elements of the
 monotone metrics for the two-level quantum systems
(cf. \cite{petzsud}), we refer the reader
to \cite{kratt}. For a further
 application --- this one to spin-1 systems --- in
which the comparative properties of priors based on the minimal and maximal
monotone metrics are assessed (with the maximal one again displaying
a certain superiority --- greater
computational tractability), see \cite{slater8}.

\acknowledgments
I would like to express appreciation to the Institute for Theoretical
Physics for computational support in this research.

\vfil
\end{document}